\documentclass[aps,prb,twocolumn,showpacs]{revtex4}

\usepackage{graphicx}
\usepackage{amsmath}
\usepackage{amssymb}

\begin{document}

\title{ARPES autocorrelation in electron-doped cuprate superconductors}

\author{Shuning Tan and Yingping Mou}
\email{yingpingmou@mail.bnu.edu.cn}

\affiliation{Department of Physics, Beijing Normal University, Beijing 100875, China}

\author{Yiqun Liu and Shiping Feng}
\email{spfeng@bnu.edu.cn}

\affiliation{Department of Physics, Beijing Normal University, Beijing 100875, China~~}

\begin{abstract}
The angle-resolved photoemission spectroscopy (ARPES) autocorrelation in the electron-doped cuprate superconductors is studied based on the kinetic-energy driven superconducting (SC) mechanism. It is shown that the strong electron correlation induces the electron Fermi surface (EFS) reconstruction, where the most of the quasiparticles locate at around the hot spots on EFS, and then these hot spots connected by the scattering wave vectors ${\bf q}_{i}$ construct an {\it octet} scattering model. In a striking analogy to the hole-doped case, the sharp ARPES autocorrelation peaks are directly correlated with the scattering wave vectors ${\bf q}_{i}$, and are weakly dispersive in momentum space. However, in a clear contrast to the hole-doped counterparts, the position of the ARPES autocorrelation peaks move toward to the opposite direction with the increase of doping. The theory also indicates that there is an intrinsic connection between the ARPES autocorrelation and quasiparticle scattering interference (QSI) in the electron-doped cuprate superconductors.
\end{abstract}


\maketitle

The undoped parent compounds of cuprate superconductors are Mott insulators. This Mott insulating-state appears to be due to the strong electron correlation \cite{Anderson87}, and yields the unconventional form of superconductivity \cite{Bednorz86,Tokura89} and anomalous normal-state properties with holes or electrons doping \cite{Timusk99}. After intensive investigations over more than 30 years, it has become clear that although the cause of the SC mechanism and the anomalous normal-state are most likely the same for both the hole- and electron-doped cases, some qualitative differences between the hole- and electron-doped cases have been observed experimentally \cite{Greene19,Armitage10}. In this case, the investigation of these differences between the hole- and electron-doped cuprate superconductors would be crucial to the understanding of the essential physics of cuprate superconductors.

Experimentally, the ARPES observations indicate that the EFS reconstruction is one of the common feature \cite{Meng09,Santander-Syro11,Korshunov05} for all families of cuprate superconductors, where the spectral weight of the quasiparticle excitation at around the antinodal regime is suppressed, and then EFS is broken up into the disconnected Fermi arcs located at around the nodal regime. However, the most of the spectral weight locates at around the tips of the Fermi arcs \cite{Vishik09,Sassa11,Horio16,Neto15,Neto16,Song17}, which in this case coincide with the hot spots on EFS. These hot spots connected by the scattering wave vector ${\bf q}_{i}$ dominate the quasiparticle scattering processes. More specifically, the ARPES data of the hole-doped counterparts show that the ARPES autocorrelation peaks that are associated with the regions of the highest joint density of states are directly correlated with these scattering wave vectors ${\bf q}_{i}$, and are well consistent with these QSI peaks observed from the Fourier transform (FT) scanning tunneling spectroscopy (STS) experiments. In this case, a natural question is whether the intrinsic feature of the ARPES autocorrelation is a universal behavior for all families of cuprate superconductors or not?

In the recent work \cite{Gao19} based on the kinetic-energy driven SC mechanism \cite{Feng0306,Feng12,Feng15a}, the ARPES autocorrelation in the hole-doped case has been discussed, and then the main features observed from the ARPES experiments \cite{Chatterjee06} are qualitatively reproduced. In this paper, we study the ARPES autocorrelation in the electron-doped side along with this line. We show explicitly that the ARPES autocorrelation peaks connected directly with the scattering wave vectors ${\bf q}_{i}$ are a universal behavior for both the hole- and electron doped cases. However, in a clear contrast to the hole-doped counterparts, the position of the ARPES autocorrelation peaks move toward to the opposite direction with the increase of doping. Our results also predict that as in the hole-doped case, there is also an intrinsic connection between the ARPES autocorrelation and QSI in the electron-doped cuprate superconductors.

The single common feature for all families of cuprate superconductors is the presence of the CuO$_{2}$ planes \cite{Timusk99,Greene19,Armitage10}, and then the experimental evidences indicate that the unconventional physics of cuprate superconductors are dominated by the CuO$_{2}$ plane. Moreover, it is widely accepted that the minimal model that may capture essential property of the CuO$_{2}$ plane is the $t$-$J$ model on a square lattice \cite{Anderson87}: $H=\sum_{l\hat{a}\sigma}t_{\hat{a}}C^{\dagger}_{l\sigma} C_{l+\hat{a}\sigma}+ \mu\sum_{l\sigma}C^{\dagger}_{l\sigma} C_{l\sigma}+J\sum_{l\hat{\eta}}S_{l}\cdot S_{l+\hat{\eta}}$, where the summation is over all sites $l$, and for each $l$, over its nearest-neighbor (NN) sites $\hat{a}=\hat{\eta}$ with the transfer integral $t_{\hat{a}}=t_{\hat{\eta}}=t$ or next NN sites $\hat{a}=\hat{\tau}$ with the transfer integral $t_{\hat{a}}=t_{\hat{\tau}}=-t'$, while the spin-exchange interaction occurs only for the NN sites $\hat{\eta}$. For the electron doping, $t<0$ and $t'<0$. $C^{\dagger}_{l\sigma}$ ($C_{l\sigma}$) is a creation (annihilation) operator for an electron with spin $\sigma$, ${\bf S}_{l}$ is the spin operator, and $\mu$ is the chemical potential. In the electron-doped case, this $t$-$J$ model is imposed a on-site local constraint $\sum_{\sigma} C^{\dagger}_{l\sigma} C_{l\sigma}\geq 1$ in order to remove zero electron occupancy of any lattice site.

In the actual analysis, the simplest realization of this local constraint is the approach based on the charge-spin separation (CSS)  \cite{Kotliar88,Yu92}. In our early studies, the CSS fermion-spin theory has been proposed for a proper treatment of no-double electron occupancy local constraint for the hole-doped case \cite{Feng9404,Feng15}. To employ this CSS fermion-spin theory to the electron-doped side, it is better to work in the hole representation via a particle-hole transformation $C_{l\sigma}\rightarrow f^{\dagger}_{l-\sigma}$, and then the $t$-$J$ model can be rewritten in the hole representation as \cite{Mou17},
\begin{eqnarray}\label{tjham1}
H=-\sum_{l\hat{a}\sigma}t_{\hat{a}}f^{\dagger}_{l+\hat{a}\sigma}f_{l\sigma}-\mu\sum_{l\sigma}f^{\dagger}_{l\sigma}f_{l\sigma}+J\sum_{l\hat{\eta}} {\bf S}_{l}\cdot {\bf S}_{l+\hat{\eta}},~~~
\end{eqnarray}
where $f^{\dagger}_{l\sigma}$ ($f_{l\sigma}$) is the creation (annihilation) operator for a hole with spin $\sigma$, and then the local constraint of no-zero electron occupancy $\sum_{\sigma}C^{\dagger}_{l\sigma}C_{l\sigma}\geq 1$ in the electron representation is replaced by the local constraint of no-double hole occupancy $\sum_{\sigma} f^{\dagger}_{l\sigma} f_{l\sigma}\leq 1$ in the hole representation. This local constraint of no-double hole occupancy now can be treat exactly within the CSS fermion-spin formalism \cite{Feng9404,Feng15}, $f_{l\uparrow}= a^{\dagger}_{l\uparrow} S^{-}_{l}$ and $f_{l\downarrow}=a^{\dagger}_{l\downarrow}S^{+}_{l}$, where the spinful fermion operator $a_{l\sigma}=e^{-i\Phi_{l\sigma}}a_{l}$ carries the charge of the constrained hole together with some effects of spin configuration rearrangements due to the presence of the doped charge carrier itself, while the spin operator $S_{l}$ describes the spin degree of freedom of the constrained hole, and then the local constraint of no-double hole occupancy is satisfied in the actual calculations. In this fermion-spin representation, the $t$-$J$ model (\ref{tjham1}) can be expressed as,
\begin{eqnarray}\label{cssham}
H&=&\sum_{l\hat{a}}t_{\hat{a}}(a^{\dagger}_{l\uparrow}a_{l+\hat{a}\uparrow}S^{+}_{l+\hat{a}}S^{-}_{l} +a^{\dagger}_{l\downarrow}a_{l+\hat{a}\downarrow}S^{-}_{l+\hat{a}}S^{+}_{l})\nonumber\\
&+&\mu\sum_{l\sigma}a^{\dagger}_{l\sigma}a_{l\sigma}+J_{{\rm eff}}\sum_{l\hat{\eta}}{\bf S}_{l}\cdot {\bf S}_{l+\hat{\eta}},
\end{eqnarray}
with $J_{{\rm eff}}=(1-\delta)^{2}J$, and $\delta=\langle a^{\dagger}_{l\sigma}a_{l\sigma}\rangle=\langle a^{\dagger}_{l}a_{l}\rangle$ that is the doping concentration. In the following discussions, the parameters in the $t$-$J$ model are chosen as $t/J=-2.5$, $t'/t=0.4$, and $J=100$ meV, which are the typical values of the electron-doped side \cite{Greene19,Armitage10}.

Within the $t$-$J$ model (\ref{cssham}) in the fermion-spin representation, the kinetic-energy driven SC mechanism has been developed \cite{Feng0306,Feng12}, where the interaction between the charge carriers directly from the kinetic energy in the $t$-$J$ model (\ref{cssham}) by the exchange of spin excitations induces the d-wave charge-carrier pairing state, then the electron pairs with the d-wave symmetry originated from the d-wave charge-carrier pairing state are due to the charge-spin recombination, and their condensation reveals the SC ground-state. Morover, we \cite{Feng15a} have developed recently a full charge-spin recombination scheme to fully recombine a charge carrier and a localized spin into a constrained electron, where the obtained electron propagator can give a consistent description of EFS both in the hole- and electron-doped cuprate superconductors \cite{Feng15a,Mou17}. Following these previous discussions \cite{Feng15a,Mou17}, the hole diagonal and off-diagonal propagators $G_{\rm f}({\bf k}, \omega)$ and $\Im^{\dagger}_{\rm f}({\bf k},\omega)$ of the $t$-$J$ model (\ref{cssham}) in the fermion-spin representation can be obtained in terms of the full charge-spin recombination scheme as,
\begin{subequations}\label{Hole-propagators}
\begin{eqnarray}
G_{\rm f}(\bf{k},\omega)&=&{1\over\omega-\varepsilon^{({\rm f})}_{\bf k}-\Sigma^{({\rm f})}_{1}({\bf k},\omega)-{[\Sigma^{({\rm f})}_{2}({\bf k}, \omega)]^{2}\over\omega+\varepsilon^{({\rm f})}_{\bf k}+\Sigma^{({\rm f})}_{1}({\bf k},-\omega)}}, ~~~~~\label{DH-propagators}\\
\Im^\dagger_{\rm f}({\bf k},\omega)&=&{-\Sigma^{({\rm f})}_{2}({\bf k},\omega)/[\omega+\varepsilon^{({\rm f})}_{\bf k}+\Sigma^{({\rm f})}_{1}({\bf k},-\omega)]\over\omega-\varepsilon^{({\rm f})}_{\bf k}-\Sigma^{({\rm f})}_{1}({\bf k},\omega)-{[\Sigma^{({\rm f})}_{2}({\bf k},\omega)]^{2}\over \omega+\varepsilon^{({\rm f})}_{\bf k}+\Sigma^{({\rm f})}_{1}({\bf k},-\omega)}}, \label{ODH-propagators}
\end{eqnarray}
\end{subequations}
where $\varepsilon^{({\rm f})}_{\bf k}=-4t\gamma_{\bf k}+4t'\gamma_{\bf k}'+\mu$ is the hole bare dispersion, with $\gamma_{\bf k}=({\rm cos} k_{x}+{\rm cos} k_{y})/2$ and $\gamma_{\bf k}'={\rm cos}k_{x}{\rm cos}k_{y}$, while the hole self-energies $\Sigma^{({\rm f})}_{1}({\bf k}, \omega)$ in the particle-hole channel and $\Sigma^{({\rm f})}_{2}({\bf k},\omega)$ in the particle-particle channel can be evaluated explicitly as,
\begin{subequations}\label{ESE1}
\begin{eqnarray}
&&\Sigma^{({\rm f})}_{1}({\bf k},\omega)={1\over N^{2}}\sum_{{\bf pp'}\nu=1,2}(-1)^{\nu+1}\Omega^{({\rm f})}_{\bf pp'k}\nonumber\\
&\times& \left [ U^{2}_{{\rm f}{\bf p}+{\bf k}} \left ( {F^{(\nu)}_{{\rm 1f}{\bf pp'k}}\over\omega+\omega_{\nu{\bf p}{\bf p}'}-E^{({\rm f})}_{{\bf p} +{\bf k}}}+{F^{(\nu)}_{{\rm 2f}{\bf pp'k}}\over\omega-\omega_{\nu{\bf p}{\bf p}'}-E^{({\rm f})}_{{\bf p}+{\bf k}}} \right ) \right. \nonumber\\
&+&\left . V^{2}_{{\rm f}{\bf p}+{\bf k}}\left ({F^{(\nu)}_{{\rm 1f}{\bf pp'k}}\over\omega-\omega_{\nu{\bf p}{\bf p}'}+E^{({\rm f})}_{{\bf p}+{\bf k} }} +{F^{(\nu)}_{{\rm 2f}{\bf pp'k}}\over\omega+\omega_{\nu{\bf p}{\bf p}'}+E^{({\rm f})}_{{\bf p}+{\bf k}}} \right ) \right ], ~~~~\label{PHESE}\\
&&\Sigma^{({\rm f})}_{2}({\bf k},\omega)={1\over N^{2}}\sum_{{\bf pp'}\nu}(-1)^{\nu}\Omega^{({\rm f})}_{\bf pp'k}{\bar{\Delta}^{({\rm f})}_{\rm Z} ({\bf p}+{\bf k})\over 2E^{({\rm f})}_{{\bf p}+ {\bf k}}} \nonumber\\
&\times& \left [\left ({F^{(\nu)}_{{\rm 1f}{\bf pp'k}}\over\omega+\omega_{\nu{\bf p}{\bf p}'}-E^{({\rm f})}_{{\bf p}+{\bf k}}}+{F^{(\nu)}_{{\rm 2f} {\bf pp'k}}\over\omega-\omega_{\nu{\bf p}{\bf p}'}-E^{({\rm f})}_{{\bf p}+{\bf k} }} \right )\right .\nonumber\\
&-&\left . \left ({F^{(\nu)}_{{\rm 1f}{\bf pp'k}}\over\omega-\omega_{\nu{\bf p}{\bf p}'}+E^{({\rm f})}_{{\bf p}+{\bf k}}}+{F^{(\nu)}_{{\rm 2f}{\bf pp'k}}\over\omega+\omega_{\nu{\bf p}{\bf p}'}+E^{({\rm f})}_{{\bf p}+{\bf k}}}\right )\right ], \label{PPESE}
\end{eqnarray}
\end{subequations}
respectively, where $U^{2}_{{\rm f}{\bf k}}=(1+ \bar{\varepsilon}^{({\rm f})}_{\bf k}/E^{({\rm f})}_{\bf k})/2$, $V^{2}_{{\rm f}{\bf k}}= (1- \bar{\varepsilon}^{({\rm f})}_{\bf k}/E^{({\rm f})}_{\bf k})/2$, $\Omega^{({\rm f})}_{\bf pp'k}=Z^{({\rm f})}_{\rm F}\Lambda^{2}_{{\bf p}+{\bf p}' +{\bf k}}B_{{\bf p}'}B_{{\bf p}+{\bf p}'}/ (4\omega_{{\bf p}'}\omega_{{\bf p}+{\bf p}'})$, $\Lambda_{\bf k}=4t\gamma_{\bf k}-4t'\gamma_{\bf k}'$, $\omega_{\nu{\bf p}{\bf p}'}=\omega_{{\bf p} +{\bf p}'}-(-1)^{\nu}\omega_{\bf p'}$, $\bar{\varepsilon}^{({\rm f})}_{\bf k}=Z^{({\rm f})}_{\rm F} \varepsilon^{({\rm f})}_{\bf k}$, $E^{({\rm f})}_{\bf k}=\sqrt{\bar{\varepsilon}^{({\rm f})2}_{\bf k}+ \mid\bar{\Delta}^{({\rm f})}_{\rm Z}({\bf k})\mid^{2}}$, the single-particle coherent weight $Z^{({\rm f})-1}_{\rm F}=1-\Sigma^{({\rm f})}_{1{\rm o}} ({\bf k},\omega=0)|_{{\bf k}=[\pi,0]}$, with $\Sigma^{({\rm f})}_{1{\rm o}}({\bf k},\omega)$ that is the antisymmetric part of $\Sigma^{({\rm f}) }_{1} ({\bf k},\omega)$, while $\bar{\Delta}^{({\rm f})}_{\rm Z}({\bf k})=Z^{({\rm f})}_{\rm F}\bar{\Delta}^{({\rm f} )}({\bf k})$, with the nonmonotonic SC gap $\bar{\Delta}^{({\rm f})}({\bf k})$ that is directly related to the electron self-energy $\Sigma^{({\rm f})}_{2}({\bf k},\omega)$ in the static-limit approximation, and can be obtained as: $\bar{\Delta}^{({\rm f})} ({\bf k})=\Sigma^{({\rm f})}_{2}({\bf k},0)=\bar{\Delta}^{({\rm f})}[({\rm cos} k_{x}-{\rm cos} k_{y})/2-B({\rm cos}2k_{x}-{\rm cos}2k_{y})/2]$, while the functions,
\begin{eqnarray*}
F^{(\nu)}_{1{\rm f}{\bf pp'k}}&=&n_{\rm F}(E^{({\rm f})}_{{\bf p}+{\bf k}}) n^{(\nu)}_{{\rm 1B}{\bf pp'}}+ n^{(\nu)}_{{\rm 2B}{\bf pp'}},\\ F^{(\nu)}_{2{\rm f}{\bf pp'k}}&=&[1- n_{\rm F}(E^{({\rm f})}_{{\bf p}+{\bf k}})]n^{(\nu)}_{{\rm 1B}{\bf pp'}} +n^{(\nu)}_{{\rm 2B}{\bf pp'}},
\end{eqnarray*}
with $n^{(\nu)}_{{\rm 1B} {\bf pp'}}=1+n_{\rm B}(\omega_{{\bf p}'+{\bf p}})+n_{\rm B}[(-1)^{\nu+1}\omega_{\bf p'}]$, $n^{(\nu)}_{{\rm 2B}{\bf pp'}} =n_{\rm B}(\omega_{{\bf p}'+{\bf p}})n_{\rm B}[(-1)^{\nu+1}\omega_{\bf p'}]$, and $n_{\rm B}(\omega)$ and $n_{\rm F} (\omega)$ that are the boson and fermion distribution functions, respectively. The spin excitation spectrum $\omega_{\bf k}$, and function $B_{\bf k}$ have been given explicitly in Ref. \onlinecite{Cheng08}, while the single-particle coherent weight $Z^{({\rm f})}_{\rm F}$, the gap parameters $\bar{\Delta}^{({\rm f})}$ and $B$, the chemical potential, together with other order parameters have been determined self-consistently.

We now turn to evaluate the electron diagonal and off-diagonal propagators $G({\bf k}, \omega)$ and $\Im^{\dagger}({\bf k},\omega)$ of the $t$-$J$ model in the SC-state, which is directly related to the hole diagonal and off-diagonal propagators $G_{\rm f}({\bf k},\omega)$ and $\Im^{\dagger}_{\rm f}({\bf k},\omega)$ in Eq. (\ref{Hole-propagators}) in terms of the particle-hole transformation $C_{l\sigma}\rightarrow f^{\dagger}_{l-\sigma}$ as $G(l-l',t-t')=\langle\langle C_{l\sigma}(t);C^{\dagger}_{l'\sigma}(t')\rangle\rangle=\langle\langle f^{\dagger}_{l\sigma}(t);f_{l'\sigma}(t')\rangle\rangle=-G_{\rm f}(l'-l, t'-t)$ and $\Im(l-l',t-t')=\langle\langle C_{l\downarrow}(t); C_{l'\uparrow}(t')\rangle\rangle=\langle\langle f^{\dagger}_{l\uparrow}(t);f^{\dagger}_{l'\downarrow}(t')\rangle\rangle=\Im^{\dagger}_{\rm f} (l-l',t-t')$. According to the above hole diagonal and off-diagonal propagators (\ref{Hole-propagators}), the electron diagonal and off-diagonal propagators are therefore obtained as \cite{Mou17} $G({\bf k},\omega) =-G_{\rm f}({\bf k},-\omega)$ and $\Im({\bf k},\omega)=\Im^{\dagger}_{\rm f} ({\bf k}, \omega)$, respectively, with the bare electron dispersion and electron self-energies that are obtained as $\varepsilon_{\bf k}=- \varepsilon^{({\rm f})}_{\bf k}$, $\Sigma_{1}({\bf k},\omega)=- \Sigma^{({\rm f})}_{1}({\bf k},-\omega)$, and $\Sigma_{2}({\bf k}, \omega)=\Sigma^{({\rm f})}_{2}({\bf k},\omega)$, respectively, and then the electron spectral function $A({\bf k},\omega)=-2{\rm Im}G({\bf k}, \omega)$ in the SC-state now can be obtained directly as,
\begin{equation}
A({\bf k},\omega)={2\Gamma({\bf k},\omega)\over [\omega-\bar{E}({\bf k},\omega)]^{2}+\Gamma^{2}({\bf k},\omega)}, \label{spectral-function}
\end{equation}
where the quasiparticle scattering rate $\Gamma({\bf k},\omega)$ and the renormalized quasiparticle dispersion $\bar{E}({\bf k},\omega)$ can be expressed explicitly as,
\begin{subequations}
\begin{eqnarray}
&&\Gamma(\bf{k},\omega)=\left | {\rm Im}\Sigma_{1}({\bf k},\omega)\right .\nonumber\\
&-&\left . {[\Sigma_{2}({\bf k},\omega)]^{2}{\rm Im}\Sigma_{1}({\bf k},-\omega)\over [\omega+\varepsilon_{\bf k}+{\rm Re}\Sigma_{1}({\bf k}, -\omega)]^{2} +[{\rm Im}\Sigma_{1}({\bf k},-\omega)]^{2}}\right |,~~~~~~~~~ \label{scattering-rate} \\
&&\bar{E}(\bf k,\omega )=\varepsilon_{\bf k}+{\rm Re}\Sigma_{1}({\bf k},\omega)\nonumber\\
&+& {[\Sigma_{2}({\bf k},\omega)]^{2}[\omega+\varepsilon_{\bf k}+{\rm Re}\Sigma_{1}({\bf k},-\omega)]\over [\omega+\varepsilon_{k}+{\rm Re}\Sigma_{1} ({\bf k},-\omega)]^{2}+[{\rm Im}\Sigma_{1}({\bf k},-\omega)]^{2}}, ~~~~~~~~~ \label{quasiparticle-dispersion}
\end{eqnarray}
\end{subequations}
respectively. In this case, the quasiparticle excitation spectrum $I({\bf k},\omega)$ can be obtained in terms of the above electron spectral function $A({\bf k}, \omega)$ as,
\begin{eqnarray}
I({\bf k},\omega)=|M({\bf k},\omega)|^{2}n_{\rm F}(\omega) A({\bf k},\omega), \label{excitation-spectrum}
\end{eqnarray}
where $M({\bf k},\omega)$ is a dipole matrix element, and does not have any significant energy or temperature dependence \cite{Zhou18}. In this case, the magnitude of $M({\bf k},\omega)$ can be rescaled to the unit. With the help of this quasiparticle excitation spectrum, the ARPES autocorrelation in the electron-doped cuprate superconductors is obtained as \cite{Chatterjee06,Mou17},
\begin{eqnarray}\label{ACF}
{\bar C}({\bf q},\omega)&=&{1\over N}\sum_{\bf k}I({\bf k}+{\bf q},\omega)I({\bf k},\omega), ~~~~~~~
\end{eqnarray}
where the summation of momentum ${\bf k}$ is restricted within the first Brillouin zone (BZ). This ARPES autocorrelation ${\bar C}({\bf q},\omega)$ therefore describes the autocorrelation of the ARPES spectral intensities at two different momenta, separated by a momentum transfer ${\bf q}$, at a fixed energy $\omega$.

\begin{figure}[h!]
\centering
\includegraphics[scale=0.68]{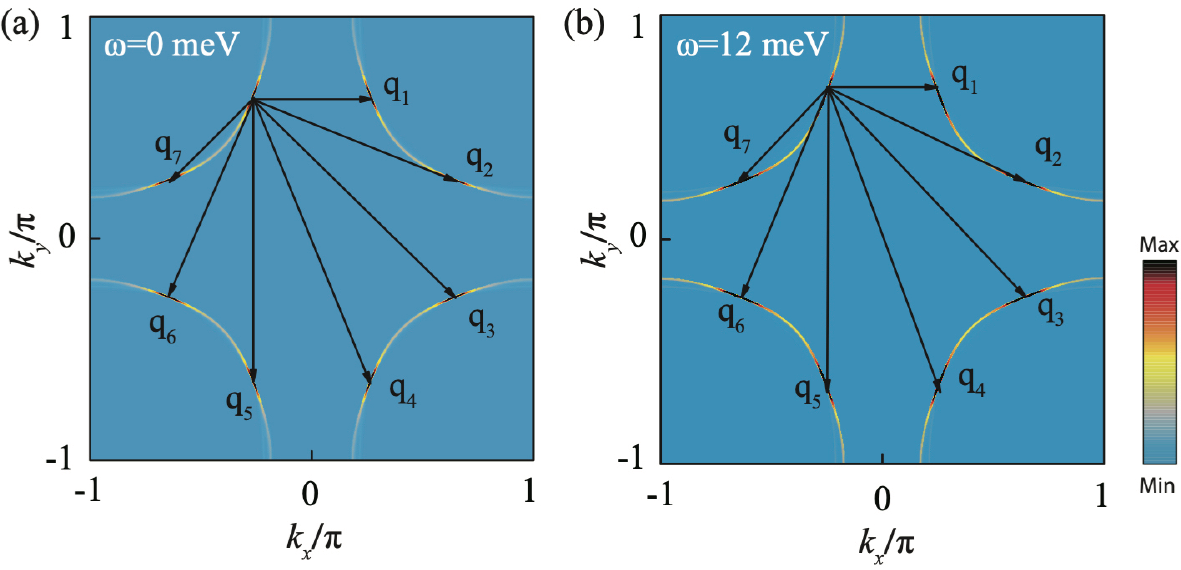}
\caption{(Color online) The maps of the quasiparticle excitation spectral intensity in (a) $\omega=0$ and (b) $\omega=12$ meV at $\delta=0.12$ with $T=0.002J$ for $t/J=-2.5$, $t'/t=0.4$, and $J=100$ meV. \label{spectral-maps}}
\end{figure}

The expression form in Eq. (\ref{ACF}) also shows that the evolution of the ARPES autocorrelation ${\bar C}({\bf q},\omega)$ with momentum, energy, and doping concentration is mainly governed by the quasiparticle excitation spectrum $I({\bf k}, \omega)$. For a complete understanding of the nature of the ARPES autocorrelation in the electron-doped cuprate superconductors, we map firstly the quasiparticle excitation spectral intensity $I({\bf k} ,\omega)$ for the binding energies (a) $\omega=0$ and (b) $\omega=12$ meV at the electron doping $\delta=0.12$ with temperature $T=0.002J$ in Fig. \ref{spectral-maps}, where as in the normal-state case \cite{Mou17}, the most exotic features can be summarized as: (a) the quasiparticle excitation spectral weight redistribution due to the strong electron correlation induces an EFS reconstruction; (b) however, the most of the spectral weight do not accumulate at around the nodes, but assembles exactly at around the tips of the Fermi arcs, which in this case coincide with the hot spots on EFS; (c) since a large part of quasiparticles meets at around the eight hot spots, the scattering wave vectors ${\bf q}_{i}$ connected with these eight hot spots form an {\it octet} scattering model; (d) these hot spots with the related {\it octet} scattering model formed in the case of the zero binding energy can persist into the case of finite binding energies (see Fig. \ref{spectral-maps}b). In particular, we \cite{Mou17} have shown that in the normal-state, the quasiparticle scattering between two hot spots on the straight Fermi arcs with the characteristic wave vector ${\bf q}_{1} ={\bf Q}_{\rm HS}$ matches well with the corresponding charge-order wave vector ${\bf Q}_{\rm CD}$ obtained in the resonant X-ray scattering measurements and STS experimental observations \cite{Horio16,Neto15,Neto16}. Incorporating the present result in the SC-state with the previous result in the normal-state  \cite{Mou17}, it is therefore shown that charge order driven by the EFS instability is developed in the normal-state \cite{Mou17}, and can persists into the SC-state, leading to a coexistence of charge order and superconductivity below $T_{\rm c}$. All these exotic features are qualitatively similar to the case occurred in the hole-doped counterparts, and are also well consistent with the experimental observations on the electron-doped cuprate superconductors \cite{Santander-Syro11,Horio16,Neto15,Neto16}. Moreover, the charge-order wave vector smoothly {\it increases} with the increase of the electron doping \cite{Mou17}, which is also consistent with the experimental results \cite{Neto16}. However, it should be emphasized that the present result of the doping dependence of the charge-order wave vector in the electron-doped case is in a striking contrast to the case in the hole-doped side, where the charge-order wave vector smoothly {\it decreases} with the increase of the hole doping \cite{Comin16,Comin14,Campi15,Peng16,Feng16}.

\begin{figure}[h!]
\centering
\includegraphics[scale=0.62]{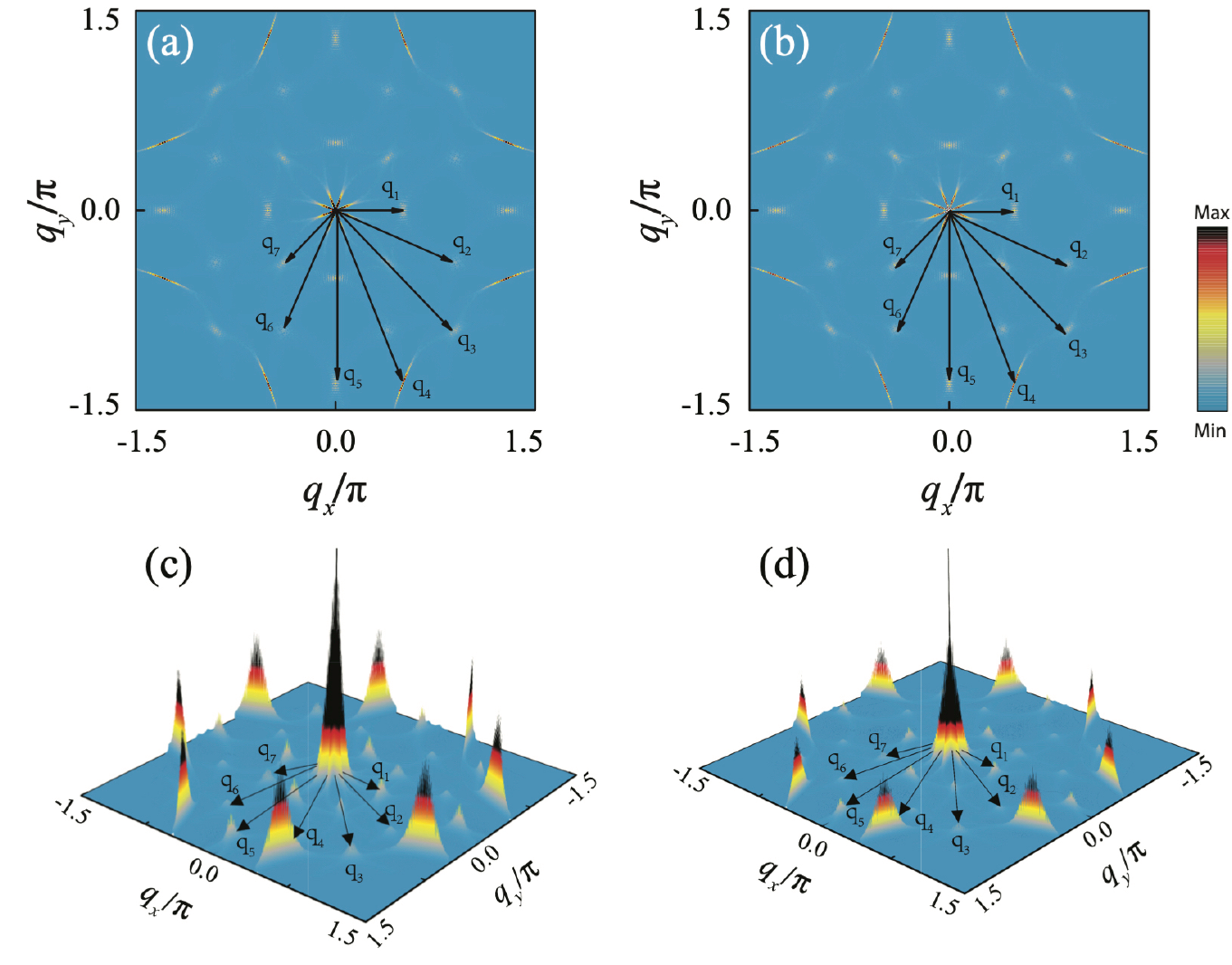}
\caption{(Color online) The maps of the intensity of the ARPES autocorrelation in (a) $\omega=12$ meV and (b) $\omega=24$ meV at $\delta=0.12$ with $T=0.002J$ for $t/J=-2.5$, $t'/t=0.4$, and $J=100$ meV. The corresponding surface plots of the ARPES autocorrelation in the $[k_{x},k_{y}]$ for (c) $\omega=12$ meV and (d) $\omega=24$ meV. \label{ARPES-autocorrelation}}
\end{figure}

\begin{figure}[h!]
\centering
\includegraphics[scale=0.76]{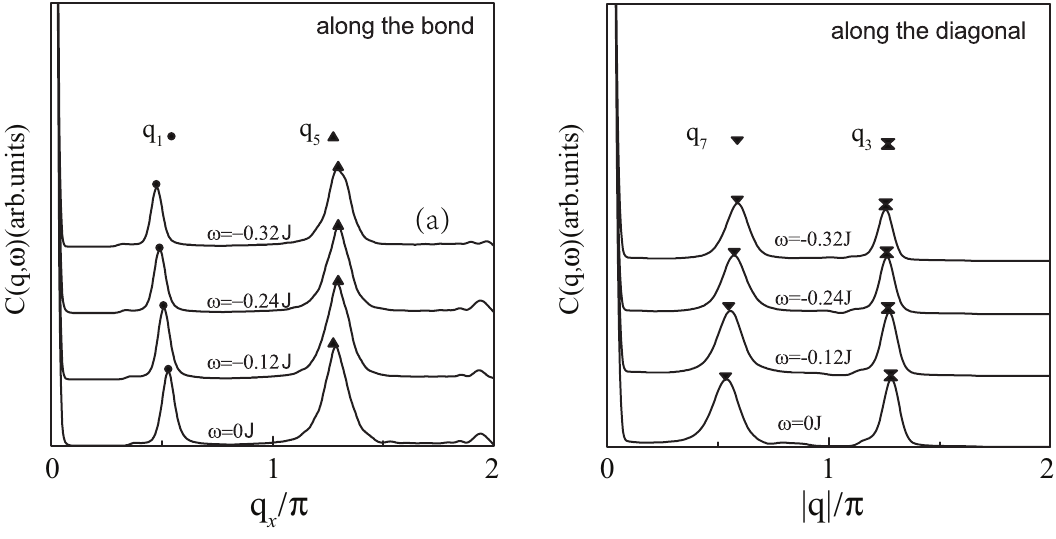}
\caption{(Color online) The ARPES autocorrelation peaks as a function of momentum with different energies along (a) the Brillouin zone parallel direction for the wave vectors ${\bf q}_{1}$ and ${\bf q}_{5}$ and (b) the Brillouin zone diagonal direction for the wave vectors ${\bf q}_{3}$ and ${\bf q}_{7}$ at $\delta=0.12$ in $T=0.002J$ for $t/J=-2.5$ and $t'/t=0.4$. \label{autocorrelation-dispersion}}
\end{figure}

We now turn to discuss the ARPES autocorrelation in the electron-doped cuprate superconductors. In Fig. \ref{ARPES-autocorrelation}, we map the SC-state ARPES autocorrelation ${\bar C}({\bf q},\omega)$ in momentum-space for the binding energies (a) $\omega=12$ meV and (b) $\omega=24$ meV at $\delta=0.12$ with $T=0.002J$. It is thus shown that some discrete spots in ${\bar C}({\bf q},\omega)$ appear, where the joint density of states is highest. To see this unusual feature more clearly, the surface plots of ${\bar C}({\bf q},\omega)$ in the $[k_{x},k_{y}]$ plane for the corresponding binding energies (c) $\omega=12$ meV and (d) $\omega=24$ meV at $\delta=0.12$ with $T=0.002J$ are shown in Fig. \ref{ARPES-autocorrelation}c and Fig. \ref{ARPES-autocorrelation}d, respectively. Obviously, the sharp ARPES autocorrelation peaks are located exactly at these corresponding discrete spots in ${\bar C}({\bf q},\omega)$. Moreover, these results in Fig. \ref{spectral-maps} and Fig. \ref{ARPES-autocorrelation} also show that the sharp ARPES autocorrelation peaks in ${\bar C}({\bf q},\omega)$ are directly correlated with these scattering wave vectors ${\bf q}_{i}$ connecting the hot spots on EFS as shown in Fig. \ref{spectral-maps}, similar to the hole-doped case \cite{Chatterjee06,Gao19}. As a nature consequence of the energy dependence of the quasiparticle excitation spectrum, the positions of the sharp peaks in ${\bar C}({\bf q},\omega)$ also vary with energy. In Fig. \ref{autocorrelation-dispersion}, we plot the positions of the sharp peaks in ${\bar C}({\bf q},\omega)$ with different energies as a function of the momentum along (a) the BZ parallel direction for the scattering wave vectors ${\bf q}_{1}$ and ${\bf q}_{5}$ and (b) the BZ diagonal direction for the scattering wave vectors ${\bf q}_{3}$ and ${\bf q}_{7}$ at $\delta=0.12$ for $T=0.002J$, where the sharp ARPES autocorrelation peaks disperse smoothly with energy, also similar to the hoel-doped case \cite{Chatterjee06,Gao19}.

\begin{figure}[h!]
\centering
\includegraphics[scale=0.76]{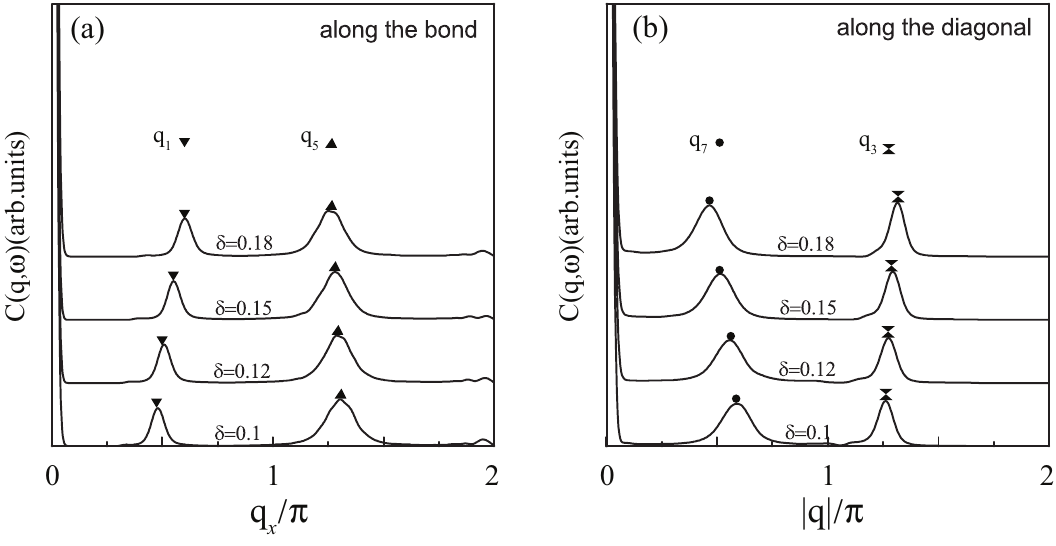}
\caption{(Color online) The ARPES autocorrelation peaks as a function of momentum at different doping concentrations along (a) the Brillouin zone parallel direction for the wave vectors ${\bf q}_{1}$ and ${\bf q}_{5}$ and (b) the Brillouin zone diagonal direction for the wave vectors ${\bf q}_{3}$ and ${\bf q}_{7}$ in $\omega=12$ meV with $T=0.002J$ for $t/J=-2.5$, $t'/t=0.4$, and $J=100$ meV. \label{autocorrelation-doping}}
\end{figure}

On the other hand, these sharp peaks in ${\bar C}({\bf q},\omega)$ are also doping dependent. For a better understanding of the evolution of the sharp peaks with doping, we plot the positions of the sharp peaks in ${\bar C}({\bf q},\omega)$ at different doping concentrations as a function of the momentum along (a) the BZ parallel direction for the scattering wave vectors ${\bf q}_{1}$ and ${\bf q}_{5}$ and (b) the BZ diagonal direction for the scattering wave vectors ${\bf q}_{3}$ and ${\bf q}_{7}$ in $\omega=12$ meV with $T=0.002J$ in Fig. \ref{autocorrelation-doping}. However, it is surprised that in a clear contrast to the hole-doped counterparts \cite{Chatterjee06,Gao19}, the positions of the sharp ARPES autocorrelation peaks with the scattering wave vectors ${\bf q}_{1}$, ${\bf q}_{3}$, ${\bf q}_{5}$, and ${\bf q}_{7}$ move toward to the opposite directions with the increase of the electron doping.

\begin{figure}[h!]
\centering
\includegraphics[scale=0.66]{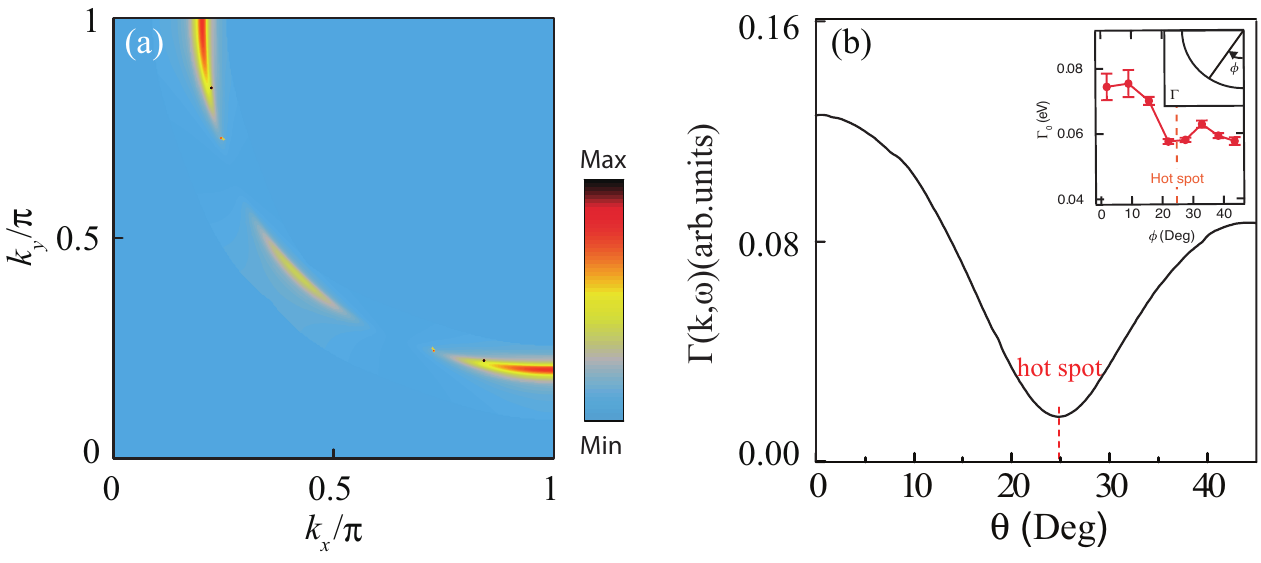}
\caption{(Color online) (a) The map of the intensity of the qusiparticle scattering rate and (b) the angular dependence of the quasiparticle scattering rate on the constant energy contour in $\omega=12$ meV at $\delta=0.12$ with $T=0.002J$ for $t/J=-2.5$, $t'/t=0.4$, and $J=100$ meV. Inset in (b): the corresponding experimental result of Pr$_{1.3-x}$La$_{0.7}$Ce$_{x}$CuO$_{4}$ taken from Ref. \onlinecite{Horio16} \label{scattering-rate}}
\end{figure}

The essential physics of the EFS reconstruction and the related {\it octet} scattering model with the scattering wave vectors ${\bf q}_{i}$ connecting the hot spots shown in Fig. \ref{spectral-maps} is the same as in the case of the normal-state \cite{Mou17}, and can be also attributed to the momentum-dependence of the quasiparticle scattering rate $\Gamma({\bf k},\omega)$ in Eq. (\ref{scattering-rate}). This follows a fact that the constant energy contour in Fig. \ref{spectral-maps} is determined by $\omega-\bar{E}({\bf k},\omega)=0$, and then the quasiparticle excitation spectral weight on the constant energy contour is dominated by the inverse of $\Gamma({\bf k},\omega)$. In Fig. \ref{scattering-rate}, we plot (a) the map of the intensity of $\Gamma({\bf k},\omega)$ and (b) the angular dependence of $\Gamma({\bf k}, \omega)$ on the constant energy contour shown in Fig. \ref{spectral-maps}b for the binding energy $\omega=12$ meV at $\delta=0.12$ with $T=0.002J$ in comparison with the corresponding experimental result  \cite{Horio16} of the angular dependence of $\Gamma({\bf k},\omega)$ observed on the underdoped Pr$_{1.3-x}$La$_{0.7}$Ce$_{x}$CuO$_{4}$ (inset in b). It thus shows that as in the normal-state case \cite{Mou17}, the minimum of $\Gamma({\bf k},\omega)$ appears exactly at the hot spots. On the other hand, the maximum of $\Gamma({\bf k} ,\omega)$ still occurs at the antinode, and then it decreases smoothly with the shift of the momentum from the antinode to the hot spot. In particular, the value of $\Gamma({\bf k},\omega)$ at the antinode is always larger than that at the node. In this case, the suppression of the quasiparticle excitation spectral weight from $\Gamma({\bf k},\omega)$ at the antinodal region is much severer than that at the nodal region. However, $\Gamma({\bf k},\omega)$ reduces lightly the quasiparticle excitation spectral weight at the hot spot region. This angular dependence of the suppression of the the quasiparticle excitation spectral weight along with the constant energy contour therefore leads to that the most of the quasiparticles are accommodated at eight hot spots, and then these eight hot spots connected by the scattering wave vectors ${\bf q}_{i}$ construct an octet scattering model as shown in Fig. \ref{spectral-maps}. This octet scattering model in turn leads to that the sharp ARPES autocorrelation peaks with the scattering wave vectors ${\bf q}_{i}$ are directly correlated to the regions of the highest joint density of states.

\begin{figure}[h!]
\centering
\includegraphics[scale=0.68]{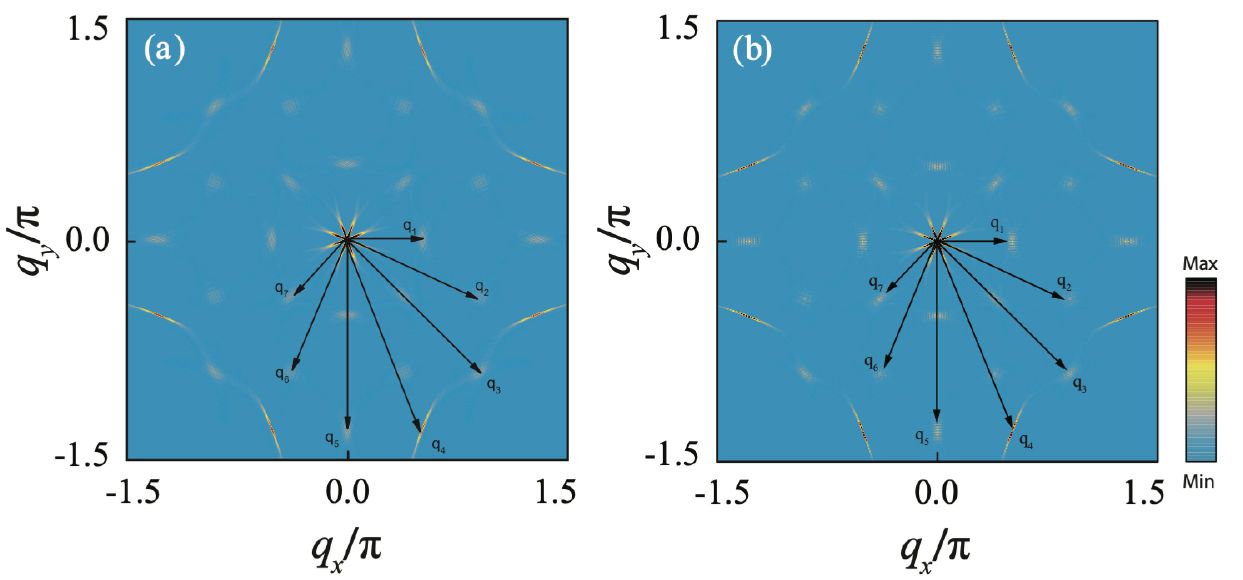}
\caption{(Color online) The maps of the intensity of (a) the Fourier transformed local density of states in the presence of a strong single point-like potential scatterer of strength $V_{0}=10$ meV and (b) the ARPES autocorrelation in $\omega=12$ meV at $\delta=0.15$ with $T=0.002J$ for $t/J=-2.5$, $t'/t=0.4$, and $J=100$ meV. \label{quaiparticle-scattering-inference}}
\end{figure}

As in the hole-doped case \cite{Chatterjee06,Gao19}, there is also an intimate connection between the ARPES autocorrelation and QSI in the electron-doped side. Theoretically, the QSI experiments in the hole-doped case can be interpreted in terms of the phenomenological {\it octet} scattering model by considering the scattering arising from a single point-like impurity \cite{Wang03,Zhang03}. In particular, the inhomogeneous part $\delta\rho({\bf q},\omega)$ of the Fourier transform (FT) local density of states for the hole-doped counterparts in the presence of a single point-like impurity scattering potential $\tilde{V}=V_{0}\delta({\bf r})\tau_{3}$ has been evaluated based on the kinetic-energy-driven SC mechanism \cite{Gao19}, and then the main features of QSI are qualitatively reproduced. Following these previous discussions \cite{Gao19}, $\delta\rho({\bf q},\omega)$ for the electron-doped side in the presence of a single point-like impurity scattering potential $\tilde{V}=V_{0}\delta({\bf r})\tau_{3}$ can be obtained directly in terms of the electron diagonal and off-diagonal Green's functions $G({\bf k},\omega)$ and $\Im^{\dagger}({\bf k},\omega)$. In Fig. \ref{quaiparticle-scattering-inference}a, we plot the momentum-space patterns of $\delta\rho({\bf q},\omega)$ for the binding energy $\omega=12$ meV at $\delta=0.15$ with $T=0.002J$ in the presence of a strong single point-like potential scatterer of the strengths $V_{0}=10$ meV. For a comparison, the corresponding result of the ARPES autocorrelation for the binding energy $\omega=12$ meV at $\delta=0.15$ with $T=0.002J$ is also plotted in Fig. \ref{quaiparticle-scattering-inference}b. These results show clearly that in a striking analogy to the hole-doped case, the obtained momentum-space structure of the $\delta\rho({\bf q},\omega)$ patterns with the strong scattering potential is qualitatively consistent with the momentum-space structure of the ARPES autocorrelation  patterns.

In conclusion, based on the kinetic-energy-driven SC mechanism, we have studied the ARPES autocorrelation in the electron-doped cuprate superconductors. Our results show that the strong electron correlation leads to the EFS reconstruction, where the most of the low-energy quasiparticles located at around the hot spots on EFS, and then these hot spots connected by the scattering wave vectors ${\bf q}_{i}$ construct an octet scattering model. In a striking analogy to the hole-doped case, the sharp ARPES autocorrelation peaks are directly correlated with the scattering wave vectors ${\bf q}_{i}$. In particular, these sharp ARPES autocorrelation peaks are weakly dispersive in momentum space. However, in a clear contrast to the hole-doped counterparts, the position of the ARPES autocorrelation peaks move toward to the opposite direction with the increase of doping. The theory also predicts that the momentum-space structure of the ARPES autocorrelation patterns in the electron-doped side is qualitatively consistent with the momentum-space structure of the QSI patterns, which should be verified by future ARPES and FT-STS experiments.

Finally, it should be emphasized that as in the hole-doped case, the strong electron correlation in the electron-doped cuprate superconductors also leads to the complicated line-shape in the quasiparticle excitation spectrum \cite{Armitage10}, such as the striking peak-dip-hump structure in the quasiparticle excitation spectrum \cite{Matsui07} and the kinks in the quasiparticle dispersion \cite{Park08}. Moreover, this strong renormalization of the electrons in turn may induce the tensile strain \cite{Saini03}. These and the related issues are under investigation now.

\section*{Acknowledgements}

This work was supported by the National Key Research and Development Program of China under Grant No. 2016YFA0300304, and the National Natural Science Foundation of China under Grant Nos. 11574032, 11734002, and 11974051.

\end{document}